\documentclass[10pt]{article}

\usepackage{amsmath}
\usepackage{amssymb}

\usepackage{graphicx}

\usepackage{cite}

\usepackage{color} 

\usepackage{setspace} 
\usepackage[labelfont=bf,labelsep=period,justification=raggedright]{caption}

\makeatletter
\renewcommand{\@biblabel}[1]{\quad#1.}
\makeatother

\date{}

\begin{document}

\begin{flushleft}
{\Large
\textbf{Quantitative Analysis  of the Effective Functional Structure   in Yeast Glycolysis}
}
\\
\vspace{1cm}
Jesus M. Cortes$^{1}$ and Ildefonso M. De la Fuente$^{2,3}$
\\
\vspace{1cm}
\bf{1} {DECSAI: Departamento de Ciencias de la Computacion e Inteligencia Artificial. Universidad de Granada, E-18071 Granada. Spain. E-mail: jcortes@decsai.ugr.es}\\
\bf{2} {Instituto de Parasitologia y Biomedicina Lopez-Neyra. CSIC. E-18100 Granada. Spain. E-mail: mtpmadei@ehu.es}\\
\bf{3} {Corresponding author}\\
\end{flushleft}


\begin{abstract}
Yeast glycolysis is considered the prototype of  dissipative biochemical
oscillators. In cellular conditions, under sinusoidal source of
glucose, the activity of glycolytic enzymes can display either periodic, quasiperiodic
or chaotic behavior.
 In order to
quantify the  functional connectivity for the glycolytic enzymes in
dissipative conditions we have analyzed  different catalytic patterns
using the non-linear statistical  tool of  Transfer Entropy. The data
were obtained by means of a yeast glycolytic model formed by three delay
differential equations where the enzymatic speed functions of the
irreversible stages  have been explicitly considered. These enzymatic activity functions
were previously modeled and tested experimentally by other different
groups.
In agreement with experimental
conditions, the studied time series corresponded
to a quasi-periodic route to chaos. The results of the analysis are three-fold: first, in
addition to the classical topological structure characterized by the
specific location of enzymes, substrates, products and feedback regulatory
metabolites, an effective functional structure  emerges in the
modeled glycolytic system, which  is dynamical and characterized by
notable variations of the functional interactions.  
Second, the dynamical  structure
exhibits a metabolic invariant which constrains the functional
attributes of the enzymes. Finally, in accordance with the classical
biochemical studies, our numerical analysis reveals  in a
quantitative manner that the enzyme phosphofructokinase is the
key-core of the metabolic system, behaving for all conditions as the
main source of the effective causal flows in yeast glycolysis.
\end{abstract}



\section*{Author Summary}

The understanding of the effective  functionality that governs the enzymatic self-organized processes in cellular conditions is a crucial topic in the post-genomic era. A number of measures have been proposed for the functionality and correlations  between biochemical time series. However,  functional   correlations do not imply effective   connectivity and most synchronization measures do not distinguish between causal and non-causal interactions. In recent studies, Transfer Entropy (TE) has been proposed as a rigorous, robust and self-consistent  method for the causal quantification of the functional information flow among nonlinear processes. Here, we have used  TE to establish the effective functional connectivity of yeast glycolysis under dissipative conditions. Concretely, we  have applied  this method for a quantification of how much the temporal evolution of the activity of one enzyme helps to improve the future prediction of another. In the enzymatic activities, the oscillatory patterns of the metabolic products might have  causal information which  can be appropriately read-out  by the TE. We have performed  numerical studies of yeast glycolysis under dissipative conditions and  found the emergence of a new kind of dynamical functional structure,  characterized by changing connectivity flows and a metabolic invariant that constrains the activity of the irreversible enzymes.

\newpage

\section*{Introduction}

Yeast glycolysis is one of the most studied dissipative pathways of the cell; it was the first metabolic system in which spontaneous oscillations were observed \cite{duysens1957,chance1964}, and the study of these rhythms allowed the construction of the first dynamic model where the  kinetics of an enzyme was explicitly considered \cite{goldbeter1972,goldbeter1973}. More concretely, the main instability-generating mechanism in the yeast glycolysis is based on the self-catalytic regulation of the enzyme phosphofructokinase \cite{goldbeter1972,boiteux1975,goldbeter2007}. 

Glycolysis is  the central pathway of glucose degradation which is implied  in  relevant  metabolic processes, such as   the maintenance of cellular redox states, the provision of ATP for membrane pumps and protein phosphorylation, biosynthesis, etc; and  its activity is linked  to a high variety  of important cellular processes, e.g.,  glycolysis has a long history in cancer cell biology   \cite{bagheri2006}  and cell proliferation \cite{almeidaa2010}, there is a correlation between brain aerobic glycolysis and Amyloid-$\beta$  plaque deposition which might  precede the clinical manifestations of the Alzheimer disease \cite{vlassenkoa2010}, the glycolytic inhibition abrogates epileptogenesis \cite{garriga2006}, and glycolysis is also related with  oxidative stress \cite{colussi2006} and apoptosis \cite{nika2003}.

Over the last 30 years a large number of different studies  focused on different  molecular mechanisms allowing for the emergency of self-organized glycolytic patterns \cite{termonia1981,dano1999,wolf2000,reijenga2002,madsen2005,olsen2009}. Nevertheless, despite the intense advance of the  knowledge of these metabolic structure, we still lack a quantitative description in cellular conditions of the effective functional structure and the causal effects among the enzymes. 

In this paper, to go a next step further in the understanding of the relationship between the classical topological structure and functionality we have analyzed the effective connectivity of yeast glycolysis, which in inter-enzyme interactions accounts for the influence that the activity of one enzyme has on the future of another \cite{gerstein1969,friston1994,fujita2007,mukhopadhyay2007,pahle2008}.

For this purpose, we  considered a yeast glycolytic model described by a system of three delay-differential equations in which there is an explicit consideration of the   speed functions of the three irreversible enzymes  hexokinase,  phosphofructokinase and  pyruvatekinase. These enzymatic activity functions were previously modeled and tested experimentally by other different groups \cite{viola1982,goldbeter1972,markus1980}.

 We  have obtained   time series of enzymatic activity under different sources  of the glucose input flux.  The data corresponded to a typical quasi-periodic route to chaos which is in agreement with experimental conditions \cite{delafuente1996a}. The dynamics of the glycolytic system changes substantially trough this route, which  allows for  a better comparison of the enzymatic processes in periodic, quasi-periodic and chaotic conditions.

Using the non-linear analysis techniques such as Transfer Entropy \cite{schreiber2000}  and Mutual Information  \cite{cover1991}, we have analyzed the glycolytic series and quantified the effective connectivity of the enzymes. 

The results show  that in the numerical analysis of yeast glycolysis, under dissipative conditions,   a effective functional structure emerges which is  characterized by changing connectivity flows and  a metabolic invariant that constraints the activity of the irreversible enzymes.


\section*{Results}

The monitoring of the fluorescence of NADH in glycolyzing baker$'$s yeast under sinusoidal glucose input flux, have shown that quasi-periodic time patterns are common at low amplitudes of the input and for high amplitudes chaotic behaviours emerge \cite{markus1985a,markus1985b}. 

In order to simulate these metabolic processes, the system is considered under periodic input flux with a sinusoidal source of glucose $\mathrm{S}=\mathrm{S_0}+ \mathrm{A} \sin (\omega t)$. Assuming the experimental value of  $\mathrm{S_0}=6$mM/h \cite{markus1984}, after dividing by $\mathrm{K}_{\mathrm{m}2}$ (the Michaelis constant of phosphofructokinase, see for more details Materials and Methods) we have obtained the normalized input flux $\mathrm{S_0}=0.033$ Hz.

Under these conditions, a wide range of different types of dynamic patterns can emerge as a function of the  control parameter, hereafter the amplitude A of the sinusoidal glucose input flux \cite{delafuente1996a,delafuente1999,delafuente1996b}.   In particular, it is observed a quasi-periodic route to chaos (cf. left panel in Fig. \ref{fig2}); thus for A$=0.001$ the biochemical oscillator exhibits a periodic pattern (Figure 2a). An increment of the amplitude to A$=0.005$ provokes a Hopf bifurcation generating another fundamental frequency, as a consequence,  quasi-periodic behaviors emerge (Figure 2b). Above A$=0.021$,  complex quasi-periodic oscillations appear (Figure 2c). After a new Hopf bifurcation  the originated dynamical behavior  is not particularly stable and small perturbations   produce  deterministic chaos (A$=0.023$, Figure 2d), as predicted by  Ruelle and Takens \cite{ruelle1971}. This route is in agreement with experimental conditions \cite{delafuente1996a}.

To go a next step further in the understanding of the relationship between the classical topological structure and effective functionality we have analyzed by means of non-linear statistical tools the catalytic patterns belonging to this scenario to chaos, and for each transition represented in the Figure 2 we have obtained three time series corresponding to the variables $\alpha$, $\beta$ and $\gamma$ (12 in total), which denote respectively the normalized concentrations of glucose-6-phosphate, fructose 1-6-bisphosphate and pyruvate.

\subsection*{Effective functionality}

Transfer Entropy (TE) quantifies the reduction in   uncertainty that one variable  has on its own future when adding another. This measure allows for a calculation of the functional influence in terms of effective connectivity between  two variables \cite{schreiber2000}. 
The analysis of the glycolytic data by means of the TE method  are shown in Table I.  The 4D vectors in square brackets  correspond to the results obtained for the 4  different amplitudes of the considered glucose input flux,  A=[0.001;0.005;0.021;0.023]. 

 The values  of functional influence  are ranging in $0.58\leq TE\leq 1.00$, with     mean$=0.79$ and   standard deviation=$0.12$,  what indicates in general terms a high  effective connectivity in the enzymatic system.
The minimum value 0.58 corresponded to the causality flow  between  E$_{3}$ and E$_{2}$ when a  simple  periodic  behavior emerges. However,  the functional connectivity from E2 to E3 shows  the maximum value,  achieved in all considered conditions of the  glucose input flux.

The glycolytic effective connectivity is illustrated in
the right panel of Fig. 2. The arrows width is proportional to
the TE between pairs of enzymes. The values change trough
the quasi-periodic route to chaos, remarked from E3 to E2 by
black dashed circles, [0.58;0.84;0.61;0.66].

In all cases analyzed, the values of TE present a maximum statistical significance (pvalue=0).



\subsection*{Total Information flows and the functional invariant}

Next, we have measured the total information flow, defined as the total outward of Transfer Entropy arriving to one enzyme minus the total inward. Positive values  mean that that enzyme is a source of causality flow and negative flows  are interpreted as sinks or targets. The results of the total information flows are shown in  Table II (pvalue=0). The maximum source of total transfer information (0.41) corresponds to the E2 enzyme (phosphofructokinase) for A=0.021, when complex quasi-periodic oscillations appear in the glycolytic system.

 For all conditions the enzyme E$_{2}$ (phosphofructokinase) is the main source of effective influence and the enzyme  E$_{3}$ (pyruvatekinase)  a sink, which could be interpreted as a target from a point of view based on its  effective functionality. The enzyme  E$_{1}$ (hexokinase) is less constrained, and it has a flow  close to zero for all conditions.

The  attributed role to each enzyme, namely  E$_{2}$ the source,   E$_{3}$ the sink and   E$_{1}$ no-constrained is an invariant and preserved trough the whole route to chaos.



\subsection*{Functional Synchronization}

Time correlations allows for 
quantification about how much two time series are statistically
independent. 
According to that, we have measured the time pairwise correlations
in the enzymatic system, and the corresponding results are shown in Table III. The main finding is  that  E$_{2}$ and E$_{3}$ are highly
  synchronized (correlation=0.90,
pvalue=0) and E$_{1}$ is
 anti-synchronized with both E$_{2}$ and E$_{3}$
(respectively, correlation equals -0.65 and -0.66, pvalue=0).

 These values  of time
 correlations were almost constant trough the quasi-periodic route to chaos and established that  the activities of E$_{2}$ and E$_{3}$ are  grouped to the same function, being activated    at similar time and
 oppositely to E$_{1}$.


\subsection*{Redundancy and uncertainty reduction}

The Mutual Information (MI) quantifies how much the knowledge of one variable reduces the entropy or uncertainty of the another \cite{cover1991}.
The analysis of the glycolytic data by means of this method 
 are shown in Table IV.

The high values of MI (close to 0.50) proved a high informative redundancy between the pairs of enzymes. So, the number of bits of information transferred from one enzyme to another   is much larger than the actually needed.

The values in the principal diagonal of Table IV represent the uncertainty for each variable. We have found these values gradually descending, H(E$_{1}$)=[1.00;1.00;1.00;1.00], H(E$_{2}$)=[0.85;0.84;0.85;0.86] and
H(E$_{2}$)= [0.76;0.74;0.76;0.78], 
which is indicative of the uncertainty in the enzymatic activity patterns belonging to E$_{1}$, E$_{2}$ and E$_{3}$ is reduced monotonously for all analyzed conditions. 

The  values of MI have a maximum statistical  significance (pvalue=0).

Finally, we have computed the Mutual Information between the glucose input
fluxes and the activity patterns of the different enzymes. In all cases, the MI was equal to zero, proving that the oscillations of
the glucose  were statistical independent of the glucose-
6-phosphate, fructose 1-6-biphosphate and pyruvate,  products of the main irreversible enzymes of glycolysis.


\section*{Discussion}

In this paper we have quantified essential aspects of the effective
functional connectivity among the main glycolytic enzymes
in dissipative conditions. 

First, we have computed under different source of glucose
the causality flows in the metabolic system.  This
level of the functional influence accounts for the contribution
of each enzyme to the generation of the different  catalytic behavior and adds a directionality in the influence
interactions between enzymes. 

The results show that the flows of functional connectivity change significantly during the different metabolic transitions analyzed, exhibiting high values of transfer entropy, and 
in all considered cases, the enzyme 
phosphofructokinase (E$_{2}$) is the main source of effective causality flow; the
pyruvatekinase (E$_{3}$) is the main sink of information flow; the
hexokinase (E$_{1}$) has a quasi-zero flow, meaning that, the total
information arriving to E$_{1}$ goes out to either E$_{2}$ or E$_{3}$.

The maximum source of total transfer information  
(0.41) corresponds to the E$_{2}$  enzyme (phosphofructokinase) at the edge of chaos,
when complex quasi-periodic oscillations emerge (cf. Fig. 2).
This finding seems to be consistent with other studies which
show that when a dynamical system operates in the frontier  between order (periodic behavior) and chaos its complexity
is maximal \cite{kaufmann1991,bertschinger2004}.

The
level of influence in terms of causal interactions between the enzymes is not always the same but varies
depending on the substrate fluxes and the dynamic characteristics
emerging in the system. In addition to the glycolitic topological
structure  characterized by the
specific location of enzymes, substrates, products and regulatory
metabolites there is an functional structure of information
flows  which is dynamic and  exhibit notable variations of the causal interactions. 

Another aspect of the glycolitic functionality was observed during the quantification of the Mutual Information,  which measures how
much the uncertainty about the one enzyme is reduced by
knowing the other;    we found that the uncertainty for E$_{1}$, E$_{2}$ and E$_{3}$   monotonously decreased for all the values of  the periodic glucose input-flux.  

Second, the numerical results show that for all analyzed cases the maximum effective connectivity  corresponds to the
Transfer Entropy from E$_{2}$ to E$_{3}$, indicating the biggest   information flow    in the multi-enzyme instability-generating
 system. This is also corroborated by the measure
of correlation between the different pairs of series which
shows that E$_{2}$ and E$_{3}$ are highly correlated, or synchronized
(correlation=0.90 pvalue=0) and E$_{1}$ is anti-correlated with
both E$_{2}$ and E$_{3}$ (respectively, correlation=-0.65 pvalue=0 and
correlation=-0.66 pvalue=0). The values of time correlations
establish that the activities of E$_{2}$ and E$_{3}$  are grouped
to the same function, being activated at similar time and oppositely
to E$_{1}$.

Third, our analysis allows for a hierarchical classification
in terms of what glycolytic enzyme is improving
the future prediction of what others, and the results reveals
in a quantitative manner that the enzyme E$_{2}$ (phosphofructokinase) is the major source of
causal information and represents the key-core of glycolysis.
The second in importance is the E$_{3}$ (pyruvatekinase).

From
the biochemical point of view the E$_{2}$ (phosphofructokinase)
has been commonly considered as a major checkpoint in the
control of glycolysis \cite{serrano1989,heinisch1996}. The main reason for this generalized
belief is  that this enzyme exhibits a complex regulatory behavior that reflects its
 capacity to integrate many different signals 	\cite{stryer1995}; from a dissipative
point of view, this enzyme catalyzes a reaction very far
from equilibrium and its self-catalytic regulation it has been
considered the main instability-generating mechanism for the
emergence of oscillatory patters in glycolysis \cite{goldbeter2007}. The  functional studies presented here confirm in a quantitative manner that
the E$_{2}$   (phosphofructokinase) is the key-core of the pathway, and our results
make stronger and expand the classical biochemical studies of
glycolysis.

Forth,  the dynamics of the glycolytic system
changes substantially trough the quasi-periodic route to chaos
when the amplitude of the input-flux varies. However, the hierarchy
obtained by transfer entropy, E$_{2}$ the flow, E$_{3}$ the sink
and E$_{1}$  a quasi-zero flow, is preserved during this route and
seems to be an invariant. This functional
invariant of a metabolic process may be important for the
understanding of functional enzymatic constraints in cellular
conditions; but this issue requires other additional studies.

Finally, we want to emphasize that Transfer Entropy as a
quantitative measure of effective causal connectivity can be a very useful
tool in studies of enzymatic processes that operate far from
equilibrium conditions. Moreover, many experimental observations have shown that the oscillations in the enzymatic activity  seem to represent one of the most striking manifestations of the metabolic dynamic behaviors, of not only qualitative but also quantitative importance in cells (further  details in Appendix III).

Transfer Entropy is
able to detect the directed exchange of causality flows among
the irreversible enzymes which might allow for  a rigorous quantification
of the effective functional connectivity of many dissipative
metabolic processes in both normal and pathological cellular
conditions.

The TE method applied to our numerical studies of yeast glycolisis  shows  the emergence of a new kind of dynamical functional structure which is characterized by changing connectivity flows and a metabolic invariant that constrains the activity of the irreversible enzymes.

The understanding the effective connectivity of the
metabolic dissipative structures is crucial to address the
functional dynamics of cellular life.

\section*{Methods}
\subsection*{Model}

\begin{figure}
\includegraphics[width=11cm]{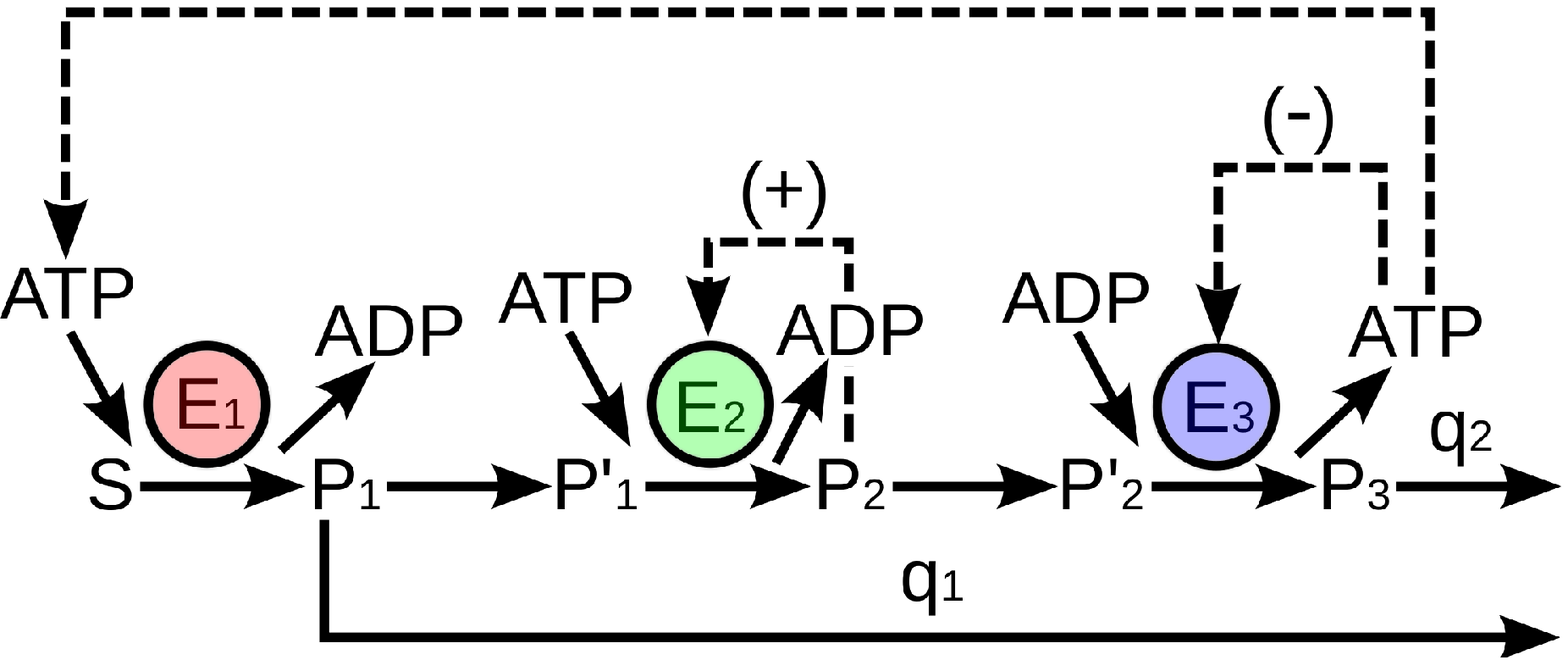}
\caption{\textbf{Multi-enzyme instability-generating system of yeast glycolysis}. The main irreversible enzymatic processes are arranged in series: E$_{1}$ (hexokinase), E$_{2}$ (phosphofructokinase) and E$_{3}$ (pyruvatekinase). S, P$_{1}$, P$'_{1}$, P$_{2}$, P$'_{2}$ and P$_{3}$ denote, respectively, the concentrations of  glucose,  glucose-6-phosphate,  fructose 6-phospfate,  fructose 1,6-bisphospfate,  phosphoenolpyruvate and   pyruvate. q$_{1}$ is the rate first-order constant for the removal of P$_{1}$;  q$_{2}$ is the rate constant for the sink of the product P$_{3}$. The model includes the feedback activation of E$_{2}$ and the feedback inhibition of E$_{3}$. The ATP is consumed by E$_{1}$ and recycled by E$_{3}$. }\label{fig1}
\end{figure}

\begin{figure}
\includegraphics[width=11cm]{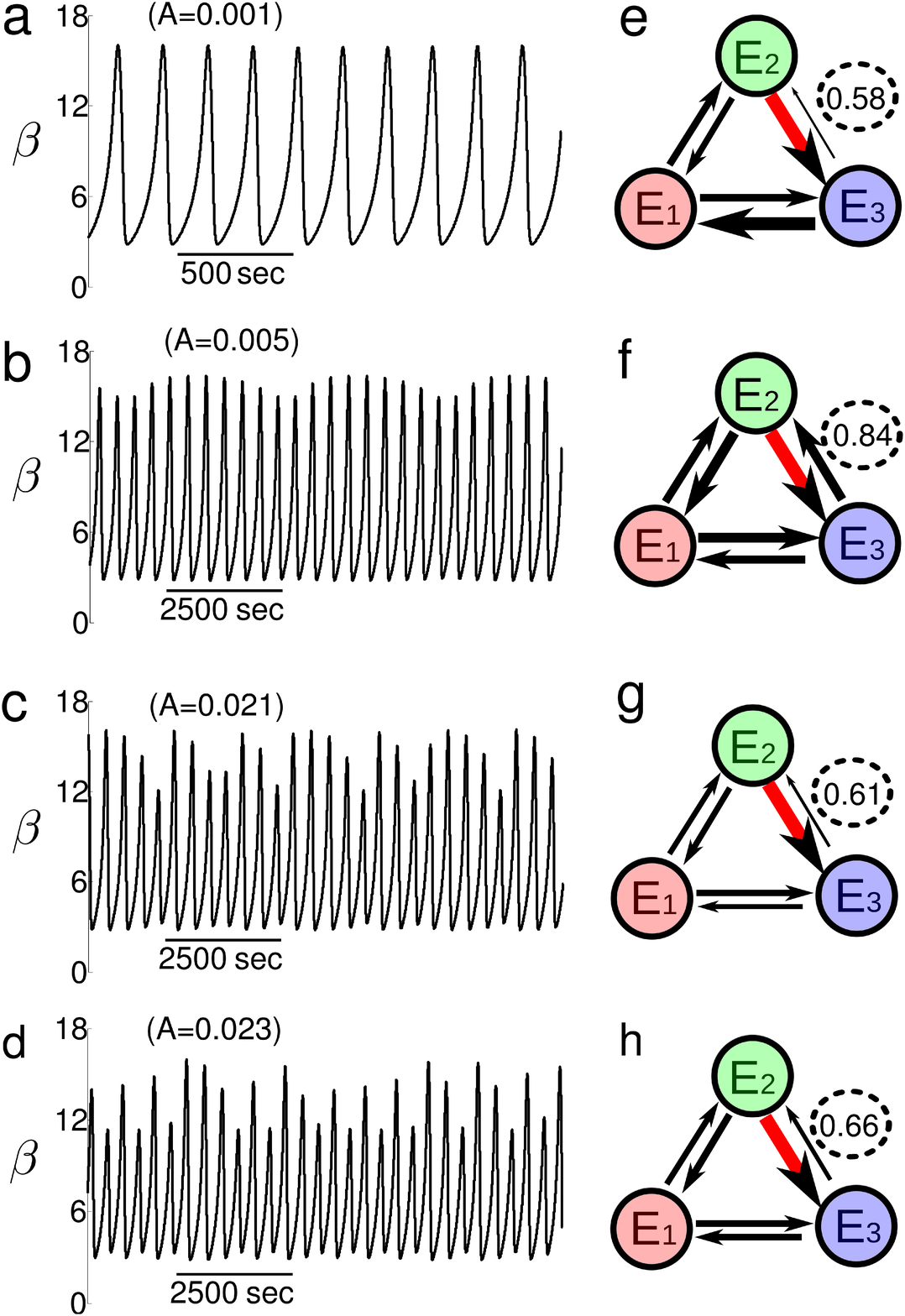}
\caption{\textbf{Glycolytic route to chaos and dynamical effective connectivity.  } Left Panel:  The time evolution of the  E$_{2}$ activity  (the normalized concentration  $\beta$, fructose 1,6-bisphospfate)  shows a quasi-periodic route to chaos  when varying the amplitude of the periodic input-flux from A$=0.001$ (top) to A$=0.023$ (bottom). (a) Periodic pattern. (b) Quasi-periodic oscillations. (c) Complex quasi-periodic motion indicating the beginning destruction of the periodic behavior. (d) Deterministic chaos.  All  series  are plotted after 10000 seconds. Right Panel: Effective connectivity of the system  for the same values of A in the left  panel. The  strength of effective connectivity is plotted with arrows  width  proportional to  the  Transfer Entropy  divided by the maximum value (red arrow), cf. results   given in Table I. Black dashed circles at the TE from E$_{3}$ and E$_{2}$ emphasize that the strength of Information flows is not the same, but varies trough the quasi-periodic route to chaos. }
\label{fig2}
\end{figure}

In Fig. \ref{fig1}   are represented the main enzymatic processes  of yeast glycolysis (the irreversible stages) with the enzymes arranged in series. 
When the metabolite S (glucose) feeds  the system,  it is transformed by the first enzyme E$_{1}$ (hexokinase) into the product P$_{1}$ (glucose-6-phosphate). The enzymes E$_{2}$ (phosphofructokinase) and E$_{3}$ (pyruvatekinase) are allosteric, and transform the substrates P$'_{1}$ (fructose 6-phosphate) and P$'_{2}$ (phosphoenolpyruvate) in the products P$_{2}$ (fructose 1-6-bisphosphate) and P$_{3}$ (pyruvate), respectively.  The step P$_2$ $\rightarrow$  P$'_2$  represents reversible activity processes, reflected in the dynamic system by the functional variable $\beta'$. 
A part of P$_{1}$ does not continue in the metabolic system, and is removed with a rate constant of q$_{1}$ which is related with the activity of pentose phosphate pathway; likewise, q$_{2}$ is the rate constant for the sink of the product P$_{3}$ which is related with the activity of pyruvate dehydrogenase complex. 

The main instability-generating mechanism in  yeast glycolysis is the self-catalytic regulation of the enzyme E$_{2}$ (phosphofructokinase), specifically, the positive feed-back exerted by the reaction products, the ADP and fructose-1,6-bisphosphate \cite{goldbeter1972,boiteux1975,goldbeter2007}.  From a strictly biochemical point of view, E$_{2}$ is also considered the main regulator enzyme of glycolysis \cite{stryer1995}.  The second irreversible stage for its regulatory importance is catalyzed by the enzyme E$_{3}$ (pyruvatekinase) which is inhibited by the ATP reaction product \cite{stryer1995}. Finally, the third irreversible process corresponds to the first stage the enzyme E$_{1}$ (hexokinase) which is dependent on the ATP.

In the determination of the enzymatic kinetics of the enzyme E$_{1}$ (hexokinase) the equation of the reaction speed dependent on glucose and ATP has been used \cite{viola1982}. The speed function of the allosteric enzyme E$_{2}$  (phosphofructokinase) was developed in the framework of the concerted transition theory \cite{goldbeter1972}. The reaction speed of the enzyme E$_{3}$ (pyruvatekinase),  dependent on ATP and phospoenolpyruvate, was also constructed on the allosteric model of the concerted transition \cite{markus1980}. 

To study the kinetics of the dissipative glycolytic system we have considered normalized concentrations;  $\alpha$, $\beta$ and   $\gamma$  denoted respectively the normalized concentrations of P$_{1}$, P$_{2}$ and P$_{3}$. For a spatially homogeneous system  the time-evolution is described by the following three delay differential equations: 
\begin{eqnarray}
\frac{\mathrm{d} \alpha}{\mathrm{d} t} &=&\mathrm{z}_1 \sigma_1 \phi_1(\mu)-\sigma_2 \phi_2(\alpha,\beta)-\mathrm{q}_1 \alpha \nonumber \\
\frac{\mathrm{d} \beta}{\mathrm{d} t} &=&\mathrm{z}_2 \sigma_2 \phi_2(\alpha,\beta)-\sigma_3 \phi_3(\beta,\beta',\mu) \nonumber \\
\frac{\mathrm{d} \gamma}{\mathrm{d} t} &=&\mathrm{z}_3 \sigma_3 \phi_3(\beta,\beta',\mu)-\mathrm{q}_2 \gamma 
\label{alpha-beta-gamma}
\end{eqnarray}
where the functional variables $\beta'$  and $\mu$ reflect the normalized concentrations of P$'_{2}$ (phosphoenolpyruvate) and ATP respectively. The three main enzymatic functions are the following:
\begin{eqnarray}
\phi_1(\mu)&=& \frac{\mu \mathrm{S}\mathrm{K}_{\mathrm{d}3}}  {\left(\mathrm{K}_{3}\mathrm{K}_{2}+\mu\mathrm{K}_{\mathrm{m}1}  \mathrm{K}_{\mathrm{d}3} + \mathrm{S}\mathrm{K}_{2}+\mu \mathrm{S}\mathrm{K}_{\mathrm{d}3}\right)}  \nonumber \\
\phi_2(\alpha,\beta)&=&\frac{\alpha \left(1+\alpha\right)\left(1+\mathrm{d}_1\beta\right)^2}{\mathrm{L}_1\left(1+\mathrm{c}\alpha\right)^2+\left(1+\alpha\right)^2\left(1+\mathrm{d}_1\beta\right)^2} \nonumber \\
\phi_3(\beta,\beta',\mu)&=&\frac{\mathrm{d}_2\beta'\left(1+\mathrm{d}_2\beta'\right)^3}{\mathrm{L}_2\left(1+\mathrm{d}_3\mu\right)^4+\left(1+\mathrm{d}_2\beta\right)^4} 
\label{phis}
\end{eqnarray}
and
\begin{eqnarray}
\beta'&=&\mathrm{f}(\beta(t-\lambda_1)) \nonumber \\
 \mu&=&\mathrm{h}(\beta(t-\lambda_2)).  
\label{fh}
\end{eqnarray} 
The constants  $\sigma_1$, $\sigma_2$ and $\sigma_3$  correspond to  the maximum activity of E$_{1}$, E$_{2}$ and E$_{3}$ ($V_{\mathrm{m}1}$, $V_{\mathrm{m}2}$ and $V_{\mathrm{m}3}$) divided by the Michaelis constants of each enzyme, respectively $\mathrm{K}_{\mathrm{m}1}$, $\mathrm{K}_{\mathrm{m}2}$ and $\mathrm{K}_{\mathrm{m}3}$. The constants z's  are defined as $\mathrm{z}_1=\mathrm{K}_{\mathrm{m}1} /\mathrm{K}_{\mathrm{m}2}$, $\mathrm{z}_2=\mathrm{K}_{\mathrm{m}2} /\mathrm{K}_{\mathrm{m}3}$ and $\mathrm{z}_3=\mathrm{K}_{\mathrm{m}3} /\mathrm{K}_{\mathrm{d}3}$, with $\mathrm{K}_{\mathrm{d}3}$ representing the dissociation constant of P$_{2}$ by E$_{3}$. The constants d's are $\mathrm{d}_1=\mathrm{K}_{\mathrm{m}3}/\mathrm{K}_{\mathrm{d}2}$, $\mathrm{d}_2=\mathrm{K}_{\mathrm{m}3}/\mathrm{K}_{\mathrm{d}3}$ and $\mathrm{d}_3=\mathrm{K}_{\mathrm{d}3}/\mathrm{K}_{\mathrm{d}4}$, with  $\mathrm{K}_{\mathrm{d}4}$ representing the dissociation constant of ATP; $\mathrm{L}_1$ and $\mathrm{L}_2$ are respectively the allosteric constant of E$_{2}$ and E$_{3}$;  c is the non-exclusive binding coefficient of the substrate P$_{1}$. More details about  parameter values and  experimental references are given in Appendix I.

From the dissipative point of view the essential enzymatic stages are those that correspond to the biochemical irreversible processes \cite{ebeling1986}  and to simplify the model, we did not consider the intermediate part of glycolysis belonging to the enzymatic reversible stages. In this way, the functions f and h are supposed to be the identity function. Thus,
\begin{eqnarray}
\beta'&= &\beta(t-\lambda_1)\nonumber \\
\mu&=&\gamma(t-\lambda_2) 
\label{betamu}
\end{eqnarray}

The initial functions present a simple harmonic oscillation in the following form:
\begin{eqnarray}
 \alpha_0(t) &=& A+B \sin(2\pi/P)\nonumber \\
\beta_0(t) &=& C+D \sin(2\pi/P)\nonumber \\ 
\gamma_0(t) &=& E+F \sin(2\pi/P)
 \label{functiont0}
\end{eqnarray}
with $A=26$, $B=12$, $C=12$, $D=10$, $E=7$, $F=6$ and $P=534$.

 The dependent variables $\alpha$, $\beta$ and   $\gamma$ were normalized dividing them by $\mathrm{K}_{\mathrm{m}2}$, $\mathrm{K}_{\mathrm{m}3}$ and $\mathrm{K}_{\mathrm{d}3}$, and the parameters $\lambda_1$  and  $\lambda_2$  are time delays affecting  the independent variable (see for more details the Appendix II).  

The numerical integration of the system  was performed with the package  ODE Workbench, which created by Dr. Aguirregabiria is part of the  Physics Academic Software. Internally this package  uses a Dormand-Prince method of order 5 to integrate differential equations.   Further information  at http://www.webassign.net/pas/ode/odewb.html.   

This model has been exhaustively analyzed before, revealing a notable richness of emergent temporal structures which included the three main routes to chaos, as well as a multiplicity of stable coexisting states, see for more details \cite{delafuente1996a,delafuente1999,delafuente1996b}.

\subsection*{Transfer Entropy} 

 TE allows  for a quantification of  how much  the temporal evolution of the activity of one enzyme  helps to improve the  future prediction of  another. The oscillatory patterns of the biochemical metabolites might have  information which can be read-out by the TE. 

For a convenient derivation, let generally assume that each of the pairs of  enzymatic activity is represented by the two  time series  $X\equiv \{x_t\}_{t=1}^T$ and $Y\equiv \{y_t\}_{t=1}^T$ . Here,  $x_t$ is the state value of the variable $X$ in time $t$, and similarly  for $y_t$. Let $I(X^P,Y^P\rightarrow X^F)=-\sum_{x_{t+1},x_t,y_t}P(x_{t+1},x_t,y_t)\log_2 P(x_{t+1}|x_t,y_t)$ be the amount of  information required to predict the future of $X$ ($X^F$)  known both the pasts of $X$ and $Y$ ($X^P$ and $Y^P$). Analogously, let  $I(X^P\rightarrow X^F)=-\sum_{x_{t+1},x_t}P(x_{t+1},x_t)\log_2 P(x_{t+1}|x_t)$ be  the amount of  information required to predict the future of $X$ known only its past. The difference $I(X^P\rightarrow X^F)-I(X^P,Y^P\rightarrow X^F)$ is by definition the transfer entropy from $Y$ to $X$, denoted by $\mathrm{TE}_{Y\rightarrow X}$. It quantifies the amount of information in digits that $Y$ adds to the predictability of $X$.  

Rewriting the   conditional probabilities  as  the joint probability divided by its marginal,  one  obtains an explicit form for the Transfer Entropy:
\begin{eqnarray}
TE_{Y\rightarrow X}=
\sum_{x_{t+1},x_t,y_t}P(x_{t+1},x_t,y_t)\log_2\left(\frac{P(x_{t+1},x_t,y_t)P(x_t)}{P(x_t,y_t)P(x_{t+1},x_t)}\right).
 \label{TE}
\end{eqnarray}

The formula (\ref{TE}) is fully equivalent to the Mutual Information between $X^F$ and $Y^P$ conditioned to $X^P$. Thus,
 $\mathrm{TE}_{Y\rightarrow X}\equiv I(X^F,Y^P|X^P)$, and consequently, Transfer Entropy says about how much information the inclusion of $Y^P$ adds to the prediction of $X^F$ only considering  $X^P$, ie. $I(X^F,Y^P|X^P)=H(X^F|X^P)-H(X^F|X^P,Y^P)$. Therefore, TE is fully quantifying the information flows between pairs of variables. The values of TE were normalized between 0 and 1.  

It is important to remark that the TE from $X$ to $Y$ is different to the one from $Y$ to $X$, ie. the effective connectivity is asymmetric, adding a  directionality in time which accounts for a particular case of \textit{directed graphs}, the graph of information flows between pairs of enzymes.

Alternatively to the Transfer Entropy, effective connectivity can be obtained using  Granger Causality \cite{granger1969}, which makes emphasis on how much  from the past of one variable the predictability of its future   is improved by adding the past of  another variable. Recently, it has been proved that in the case of Gaussian variables both  Transfer Entropy and Granger Causality are measuring exactly the same \cite{barnett2009}. Therefore, the information flows based on Transfer Entropy and the Granger causality interactions   coincide for Gaussian variables.

\section*{Mutual Information and Redundancy}

MI  quantifies how much the knowledge of one variable reduces the entropy or uncertainty of  another. Therefore, MI says about how much information the two variables are sharing. The strongest point of the MI is   that it extends  functionality to high order statistics \cite{cover1991}. 
Its definition is  $MI(X,Y)=H(X)-H(X|Y)$, where $H(X|Y)=H(X,Y)-H(Y)$ is the conditional entropy of $X$ given $Y$.  It  accounts for the remaining uncertainty in $X$ knowing the variable $Y$. We referred $H(X)$ and $H(X,Y)$ as respectively the joint and marginal (Shanon) entropies.

For statistical independent $X$ and $Y$ variables one has $MI(X,Y)=0$. The other limit satisfies  $MI(X,X)=H(X)$, because of $H(X|Y)=0$. Therefore, the MI  of two variables is bounded and    satisfies that $0\leq MI(X,Y)\leq H(X)$.  High values of MI mean that the redundancy in information between the two variables is large. The values of MI were normalized between 0 and 1.

\section*{Number of bins  vs Statistical significance}

For all the probabilities used in both Transfer Entropy (TE) and Mutual Information (MI) we used a number of bins of  10. As it is well-known, the calculation of these probabilities is sensitive to the number of bins. Instead of tuning it as a control parameter to compute the probabilities, we preferred explored the statistical significance of the computed values. This was achieved by comparing both the TE and MI  between the two series of enzymatic activity, say  X and Y, with the values obtained  when considering  a random permutation of  Y, what we called, the shuffled Y.  
The  values of both TE and MI  shown in Tables I and III
  were larger than those calculated in the shuffled situation (for both TE and MI, pvalue=0, for 50 different samples).

\section*{Acknowledgments}

JMC is funded by  the Spanish Ministerio de Ciencia e Innovacion, programa Ramon y Cajal, and from Junta de Andalucia, grants P09-FQM-4682 and P07-FQM-02725. I.M. De la Fuente acknowledges useful advises and suggestions from Prof. J. Veguillas.

\newpage


\newpage

\newpage

\newpage
\begin{figure}
\includegraphics[width=11cm]{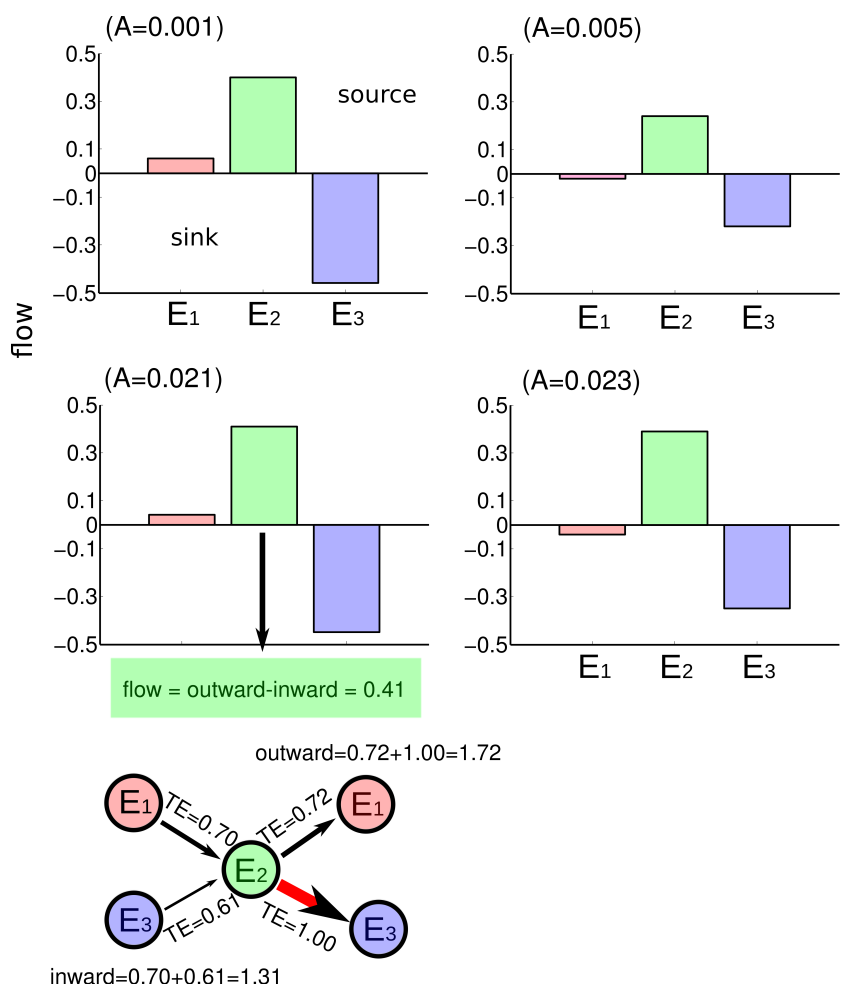}
\caption{\textbf{Total information flows and the functional invariant.} Bars represent the total information flow, defined per each enzyme as the total outward TE minus the total inward. For A=0.021 and  E$_{2}$  an schematic visualization of the calculation of this flow is shown (bottom graph of the panel). The functionality attributed for each enzyme  is an invariant and preserved along the route, ie. E$_{2}$ is a source, E$_{3}$ is a sink and E$_{1}$ has a quasi-zero flow.  }
\label{fig3}
\end{figure}

\clearpage
\newpage
\section*{Tables}
\textbf{Table I. Values of normalized Transfer Entropy}
\begin{table}[ht]
\centering 
\begin{tabular}{|c|c|c|c|} 
\hline 
  & \textbf{From  E}$_{1}$ & \textbf{From  E}$_{2}$   & \textbf{From  E}$_{3}$  \\ [0.5ex] 
\hline 
\textbf{To  E}$_{1}$ & --- & [0.73;0.88;0.72;0.76]  & [0.74;0.80;0.68;0.74] \\ 
\hline 
\textbf{To  E}$_{2}$ & [0.76;0.80;0.70;0.72] & --- & [0.58;0.84;0.61;0.66]  \\
\hline 
\textbf{To  E}$_{3}$ & [0.78; 0.86;0.74;0.75]&   [1.00;1.00;1.00;1.00]  & --- \\ [1ex] 
\hline 
\end{tabular}
\end{table}

\newpage
\textbf{Table II. Values of total information flows}
\begin{table}[ht]
\centering 
\begin{tabular}{|c|c|c|} 
\hline 
\textbf{E}$_{1}$ &[0.06; -0.02;0.04;-0.04] & Quasi-zero flow \\
\hline 
\textbf{E}$_{2}$ &[0.40; 0.24;0.41;0.39]& source \\
\hline 
\textbf{E}$_{3}$ &[-0.46; -0.22;-0.45;-0.35]& sink \\  
\hline 
\end{tabular}
\end{table}

\newpage
\vspace{0.5cm}
 \textbf{Table III. Time Correlations}
 \vspace{0.5cm}

\begin{table}[ht]
\centering 
\begin{tabular}{|c|c|c|c|} 
\hline 
  & \textbf{E}$_{1}$ & \textbf{E}$_{2}$   & \textbf{E}$_{3}$  \\ [0.5ex] 
\hline 
\textbf{E}$_{1}$ & [1.00;1.00;1.00;1.00]  & [-0.65;-0.66;-0.64;-0.63]  &  [-0.66;-0.66;-0.66;-0.66]  \\ 
\hline 
\textbf{E}$_{2}$ & [-0.65;-0.66;-0.64;-0.63]  & [1.00;1.00;1.00;1.00] & [0.90;0.90;0.90;0.90]  \\
\hline 
\textbf{E}$_{3}$ & [-0.66;-0.66;-0.66;-0.66]& [0.90;0.90;0.90;0.90]   &  [1.00;1.00;1.00;1.00]\\ [1ex] 
\hline 
\end{tabular}
\end{table}

\newpage
\vspace{0.5cm}
 \textbf{Table IV. Values of normalized Mutual Information}
 \vspace{0.5cm}

\begin{table}[ht]
\centering 
\begin{tabular}{|c|c|c|c|} 
\hline 
  & \textbf{E}$_{1}$ & \textbf{E}$_{2}$   & \textbf{E}$_{3}$  \\ [0.5ex] 
\hline 
\textbf{E}$_{1}$ & [1.00;1.00;1.00;1.00]  & [0.52;0.49;0.45;0.44]  & [0.48;0.49;0.45;0.44] \\ 
\hline 
\textbf{E}$_{2}$ & [0.52;0.49;0.45;0.44] & [0.85;0.84;0.85;0.86] & [0.47;0.46;0.45;0.45]  \\
\hline 
\textbf{E}$_{3}$ & [0.48;0.49;0.45;0.44]& [0.47;0.46;0.45;0.45]    &  [0.76;0.74;0.76;0.78]\\ [1ex] 
\hline 
\end{tabular}
\end{table}

\end{document}